\documentclass[10pt,aps,prd,amsmath,amssymb,superscriptaddress,showkeys,showpacs,twocolumn,floatfix,nofootinbib]{revtex4-1}
\usepackage{mathrsfs,amsmath,amsthm,latexsym,amssymb,amsfonts,epsfig,cancel,enumerate,graphicx,txfonts,diagbox, CJKutf8}
\usepackage[utf8]{inputenc}  
\usepackage[T1]{fontenc}     
\usepackage{lmodern}         
\usepackage[percent]{overpic}
\usepackage{multirow}
\usepackage[T1]{fontenc}
\usepackage[utf8]{inputenc}
\usepackage[colorlinks=true,linkcolor=blue,citecolor=blue,urlcolor=blue]{hyperref}

\makeatletter
\providecommand{\href@noop}[2]{#2}
\makeatother
\usepackage{orcidlink}

\usepackage{xcolor}

\usepackage{booktabs}
\usepackage{graphicx}
\definecolor{navy}{RGB}{0,0,150}
\usepackage{appendix}
\usepackage{needspace}

\allowdisplaybreaks

\newcommand{\RGUI}{Department of Physics and Chongqing Key Laboratory for Strongly Coupled Physics, Chongqing University, Chongqing 401331, P. R. China}

\begin{document}
\begin{CJK}{UTF8}{gbsn}
\baselineskip=14pt

\title{Vacuum Decay and Baryogenesis Associated with Primordial  Black Holes at Finite Temperature}
\author{Qi Sun}

\affiliation{\RGUI}
\author{Ligong Bian}
\email{lgbycl@cqu.edu.cn}
\affiliation{\RGUI}

\begin{abstract}
Within finite-temperature effective potential framework, we study the effects of PBHs on vacuum decay, the associated gravitational wave spectrum, and the baryon asymmetry. For the PBHs considered, the Hawking temperatures are below the ambient temperature, and their dominant effect is taken to be the gravitational distortion of the Higgs field configuration. Relative to flat spacetime, the PBH background reduces the vacuum decay exponent, enhances the gravitational wave peak amplitude, and shifts the peak frequency downward. For PBH masses of roughly $10^{11}~{\rm g}-10^{13}~{\rm g}$, the sphaleron energy is reduced, while the small PBH abundance renders the associated washout negligible. Additionally, over roughly the same mass range, PBH-emitted BSM particle decays generate a baryon asymmetry too small to spoil the agreement between the baryon abundances inferred from BBN and CMB.
\end{abstract}

\maketitle

\section{Introduction}
 In 2012, the key ingredient for the the Higgs mechanism \cite{englert1964broken,higgs1964broken,Guralnik:1964eu}, i.e., the 125 GeV Higgs boson was discovered by the ATLAS and CMS Collaborations \cite{aad2012combined,chatrchyan2012combined}. Since then, increasingly precise measurements of the Higgs boson mass, the top quark mass, and the strong coupling constant have enabled detailed studies of the stability of the Standard Model (SM) electroweak vacuum. For the current experimental central values of these quantities, the SM Higgs effective potential may develop an instability at large field values, suggesting that the present electroweak vacuum is metastable \cite{degrassi2012higgs,gorsky2015standard,bezrukov2015should,ellis2015discrete,krive1976vacuum,turner1982our,sher1989electroweak,isidori2001metastability}. Nevertheless, the corresponding vacuum lifetime is predicted to be much longer than the age of the Universe \cite{isidori2001metastability,Elias-Miro:2011sqh,degrassi2012higgs}.

The semiclassical theory of vacuum decay was developed by Coleman \cite{coleman1977fate} and later generalized by Coleman and De Luccia to include gravitational effects \cite{coleman1980gravitational}. The standard Coleman-De Luccia picture describes bubble nucleation in a homogeneous false vacuum background. More generally, localized inhomogeneities may serve as nucleation sites and thereby modify the vacuum decay rate \cite{hiscock1996nucleation,shukla2000inhomogeneous,kasai2015decay,oshita2019compact,koga2020instability}. In the early Universe, black holes may provide such localized gravitational seeds \cite{gregory2014black,2015Gravity,burda2016fate,tetradis2016black,canko2018catalysis,dai2020connecting,zeng2024phase}. For sufficiently small black holes, thermodynamic effects associated with the Bekenstein-Hawking entropy. This picture can be further extended by including Hawking-radiation-induced hotspots, which can affect vacuum decay when the hotspot temperature exceeds the ambient plasma temperature \cite{Hamaide:2023ayu}. For black holes with horizon scales comparable to the bubble size, geometric effects may significantly reduce the decay exponent. By modifying the nucleation process, these effects can influence the subsequent evolution of the first order phase transition and thereby alter the associated gravitational wave (GW) spectrum \cite{Zhong:2025xwm,Tanaka:2026geo}. 

In addition to vacuum decay, sphaleron transitions constitute another important class of non-perturbative processes in the electroweak theory. In the SM \cite{glashow1961partial,higgs1964broken,englert1964broken,weinberg1967model,fritzsch1973advantages}, non-perturbative electroweak gauge field configurations can induce processes that violate $B+L$ while conserving $B-L$. Black hole background might also affect the electroweak sphaleron process. A previous study showed that a black hole with a Schwarzschild radius comparable to the characteristic sphaleron size can act as a localized gravitational impurity, lowering the sphaleron energy barrier and thereby enhancing local $B+L$-violating processes \cite{de2021standard}. Black holes may affect the baryon asymmetry generation not only by catalyzing sphaleron transitions but also through Hawking evaporation. In this scenario, black holes nonthermally emit beyond Standard Model (BSM) particles $X$, whose subsequent baryon-number-violating and CP-violating decays can generate a net baryon asymmetry, provided that it is not erased by washout processes. Previous studies show that sufficiently light primordial black holes (PBHs) scenarios that evaporate before the onset of Big Bang nucleosynthesis (BBN) can reproduce the observed baryon asymmetry in certain regions of parameter space \cite{Gehrman:2022imk,Barman:2022pdo,ParticleDataGroup:2020ssz}. 

In this work, we investigate black hole assisted vacuum decay, the associated GW spectrum, and baryogenesis from PBH evaporation. In Sec.~\ref{sec:vacuum_decay}, we study the effect of a black hole background on the vacuum decay exponent within the one-loop finite-temperature effective potential and compare it with the corresponding flat spacetime case. In Sec.~\ref{sec:gw}, we discuss the resulting first order phase transition and GW spectrum. In Sec.~\ref{sec:sphaleron}, we analyze the sphaleron energy in a black hole background and its possible implications for the baryon asymmetry of the early Universe. In Sec.~\ref{sec:baryon}, we evaluate the baryon asymmetry generated by the decays of BSM particles emitted by PBHs via Hawking radiation. We summarize our conclusions in Sec.~\ref{sec:conclusion}.

\section{Black hole seeded vacuum decay in a finite-temperature effective potential}\label{sec:vacuum_decay}
In this section, we study false vacuum decay in a black hole background using finite-temperature effective potential $V(\phi,T)$ normalized with respect to the false vacuum, $V(\phi_{\rm fv},T)=0$.

\subsection{Finite-temperature effective potential}
The finite-temperature effective potential is evaluated using the background-field method in the Landau gauge and the modified minimal subtraction ($\overline{\mathrm{MS}}$) renormalization scheme. The daisy improved one-loop finite-temperature effective potential is written as
\begin{equation}
    V(\phi,T)=V_{\rm tree}+V_{\rm CW}+V_{\rm T}+V_{\rm daisy}.
\end{equation}
The tree-level contribution $V_{\rm tree}$ represents the tree-level potential, supplemented by a dimension-six operator that can generate a barrier between the symmetric and the broken phases for suitable parameter choices. The tree-level potential is given by
\begin{equation}
    V_{\rm tree}=-\frac{1}{2}m^{2}\phi^{2}+\frac{\lambda}{4} \phi^{4}+\frac{\phi^{6}}{8\Lambda^{2}},
\end{equation}
where $\Lambda$ denotes the new physics (NP) scale. Requiring the electroweak minimum to be the global minimum imposes $\Lambda \ge \frac{v^2}{m_h}$, while a potential barrier can develop for $\Lambda \le \frac{\sqrt{3}v^2}{m_h}$, thereby allowing for a first-order phase transition \cite{Grojean:2004xa,Huang:2015tdv}. The Coleman--Weinberg term $V_{\rm CW}$ describes the zero-temperature one-loop contribution \cite{Ellis:2018mja}:
\begin{equation}
    V_{\rm CW}=\sum_{i} n_{i} \frac{m_{i}^{4}}{64 \pi^{2}}\left(\log \left(\frac{m_{i}^{2}}{\bar{\mu}^{2}}\right)-c_{i}\right).
\end{equation}
Here, $i={\{ \phi, \chi_{1}, \chi_{2}, \chi_{3}, W, Z, t\}}$. The quantity $n_{i}$ denotes the degrees of freedom, with $n_{\phi, \chi_{1}, \chi_{2}, \chi_{3}, W, Z, t}=\{1,1,1,1,6,3,-12\}$. For gauge bosons, $c_{i} = \frac{5}{6}$, while for scalars and fermions, $c_{i} = \frac{3}{2}$. We set the renormalization scale $\bar{\mu}=T$ throughout our analysis. The one-loop finite-temperature contribution can be expressed as \cite{Bernon:2017jgv}
\begin{equation}
\begin{aligned}
V_{\rm T}
&=\sum_{i} n_{i} \frac{T^{4}}{2 \pi^{2}} J_{b,f} \\
&=\mp \sum_{i} n_{i} \frac{T^{4}}{2 \pi^{2}}
\left(
\sum_{l=1}^{n}
\frac{(\pm 1)^{l}}{l^{2}}
\left(\frac{m_{i}}{T}\right)^{2}
K_{2}\left(\frac{m_{i}}{T}l\right)
\right).
\end{aligned}
\end{equation}
where $K_{2}$ denotes the modified Bessel function of the second kind and $n=7$. To account for infrared divergences associated with the Matsubara zero-modes of bosons, we introduce a daisy resummed term \cite{croon2021theoretical}
\begin{equation}
    V_{\rm daisy}=-\sum_{i} n_{i} \frac{T}{12 \pi} (m_{i,res}^3- m_{i}^3),
\end{equation}
in which $i=\{ \phi, \chi_{1}, \chi_{2}, \chi_{3}, W_{L}, Z_{L}, \gamma_{L}\}$ and the subscript $L$ labels the longitudinal gauge boson. The degrees of freedom factors entering the daisy term are $n_{ \phi, \chi_{1}, \chi_{2}, \chi_{3}, W_{L}, Z_{L}, \gamma_{L}}=\{1,1,1,1,2,1,1\}$. The quantity $m_{i, res}$ denotes the resummed mass. The explicit expressions for the field-dependent masses and thermal self-energies are given in Ref.~\cite{zhu2025theoretical}. 

The resulting effective potential determines the thermal evolution of the electroweak vacuum. At sufficiently high temperatures, thermal effects restore the electroweak symmetry and the finite-temperature effective potential is minimized at $\phi=0$. When the Universe cools to the critical temperature $T_{c}$, the symmetric and broken minima become degenerate and satisfy 
\begin{align}
V(\phi=\phi_c, T_c)&=V(0, T_c), \\ \frac{\partial V(\phi, T_c)}{\partial \phi}\Big|_{\phi=\phi_c}&=0. 
\end{align}
Upon further cooling, bubbles of the broken phase can nucleate within the false vacuum. The nucleation temperature $T_{n}$ is determined by $\frac{\Gamma(T_n)}{H^4(T_n)}\sim 1$.

\subsection{Equations of motion for black hole seeded vacuum decay}
Assume that a black hole is present in the false vacuum. The black hole acts as a localized gravitational inhomogeneity and may serve as a nucleation site for a critical bubble. To describe this process, we start from the Lorentzian action for a scalar field minimally coupled to gravity \cite{canko2018catalysis},
\begin{align}
    S&=\int d^{4}x \sqrt{-g} \left[ -\frac{1}{2}g^{ \mu \nu} \partial_{ \mu} \phi \partial_{ \nu} \phi-V \left( \phi,T \right)+ \left( 16 \pi G \right)^{-1}\mathcal{R} \right] \notag\\
    &+\frac{1}{8\pi G}\int_{\partial\mathcal{M}}K\sqrt{\gamma}\mathrm{d}^{3}y,
\end{align}
where the last term is the Gibbons-Hawking-York boundary term required for a well-defined gravitational variational principle. We assume that the spacetime geometry is static and $O(3)$ symmetric. Accordingly, we adopt the following Schwarzschild-like ansatz in Lorentzian signature,
\begin{align}
   \mathrm{d}s^{2}&=-N(r) e^{2 \delta(r)}\mathrm{d} t^{2}+N^{-1}(r) \mathrm{d}r^{2} \notag  \\
   &+ r^{2}(\mathrm{d}\theta^{2}+\sin^{2}\theta\mathrm{d}\varphi^{2 }),\\ N(r)&=1-\frac{2GM(r)}{r}.
   \label{sch1}
\end{align}
The event horizon ($r=R_h$) is determined by $N(R_h)=0$, which gives $M(R_h)=R_h/(2G) \equiv M_{\rm BH}$, with $M_{\rm BH}$ denoting the mass of the seed black hole.

We consider a static critical bubble centered on the seed black hole. Neglecting the backreaction of the scalar configuration on the geometry, we approximate the background spacetime by the Schwarzschild metric. Since the gravitational background is dominated by the seed black hole, we write $M(r)=\frac{R_h}{2G}+\delta M(r)$ and keep the leading contributions in $G$. Moreover, we set $\delta'=0$ and $\delta=0$. Thus, the scalar field equation and the corresponding equation for the mass correction are
\begin{align}
\phi^{\prime\prime}&=\frac{r}{r-R_{\mathrm{h}}}\frac{dV(\phi,T)}{d\phi}-\left(\frac{2}{r}+\frac{R_{\mathrm{h}}}{r(r-R_{\mathrm{h}})}\right)\phi^{\prime},\label{field_equation1}\\
\delta M^{\prime}&=4\pi r^{2}\left[\frac{1}{2}\frac{r-R_{\mathrm{h}}}{r}\phi^{\prime 2}+V(\phi,T)\right],
\end{align} where primes denote derivatives with respect to $r$. At spatial infinity, we impose $\phi(r)\to\phi_{\rm fv}$ as $r\to\infty$, so that the scalar configuration approaches the false vacuum and its contribution to the ADM mass remains finite. Regularity of the field configuration at the horizon requires 
\begin{equation}
    \phi^{\prime}(R_h)=R_h \frac{ \partial V(\phi(R_h),T)}{\partial\phi}. 
    \label{boundary}
\end{equation}
In the limit $R_h\to0$, this condition reduces to the boundary condition in flat spacetime, $\phi^{\prime}(0)=0$. For a solution satisfying the above boundary conditions, the total mass correction associated with the scalar configuration is therefore
\begin{equation}
\delta M_{\rm tot}(R_h,T)=4 \pi \int_{R_{h}}^{ \infty}dr r^{2}
\left[ \frac{1}{2} \frac{r-R_{h}}{r}\phi^{\prime 2}
+V \left( \phi,T \right) \right].
\end{equation}
The quantity $\delta M_{\rm tot}(R_h,T)$ represents the energy barrier associated with the critical bubble in the black hole background approximation.

\subsection{Decay exponent in flat spacetime and black hole backgrounds}
The decay exponent in a black hole background is given by
\begin{equation}
B(R_h,T)=\frac{\delta M_{\rm tot}(R_h,T)}{T}.
\end{equation}
Following Ref.~\cite{canko2018catalysis}, we consider the regime $T_{\rm BH}<T$ and neglect the influence of Hawking radiation, where $T_{\rm BH}=M_{\rm Pl}^2/(8\pi M_{\rm BH})$ is the Hawking temperature of the black hole. The temperature $T$ entering the decay exponent therefore denotes the ambient temperature. In this approximation, the dominant effect of black hole is its gravitational modification of the field configuration.

In the flat spacetime limit, $R_h\to 0$, the decay exponent reduces to its flat spacetime form,
\begin{align}
B_{\rm flat}(T)&\equiv \lim_{R_h\to 0}B(R_h,T) \notag\\
&=\frac{\displaystyle\lim_{R_h\to 0}\delta M_{\rm tot}(R_h,T)}{T}=\frac{S_3(T)}{T}.
\end{align}
For an electroweak scale phase transition in a radiation-dominated Universe, the above nucleation condition $\frac{\Gamma(T_n)}{H^4(T_n)}\sim 1$ is commonly approximated by $\frac{S_{3}(T)}{T} \simeq 140$. For a static $O(3)$ symmetric configuration at finite temperature, $S_3(T)$ is three-dimensional Euclidean action,
\begin{equation}
S_{3}(T)
=
4\pi\int_{0}^{\infty}dr\, r^{2}
\left[
\frac{1}{2}\left(\frac{d\phi}{dr}\right)^{2}
+V(\phi,T)\right].
\end{equation}
The corresponding bounce equation is
\begin{equation}
\frac{d^{2}\phi}{dr^{2}}
+\frac{2}{r}\frac{d\phi}{dr}
=
\frac{\partial V(\phi,T)}{\partial\phi},
\end{equation}
with boundary conditions
\begin{equation}
\left.\frac{d\phi}{dr}\right|_{r=0}=0,
\qquad
\phi(r\to\infty)=\phi_{\rm fv}.
\end{equation}

\subsection{Numerical results}
\begin{figure}[!htp]
\includegraphics[width=0.47\textwidth]{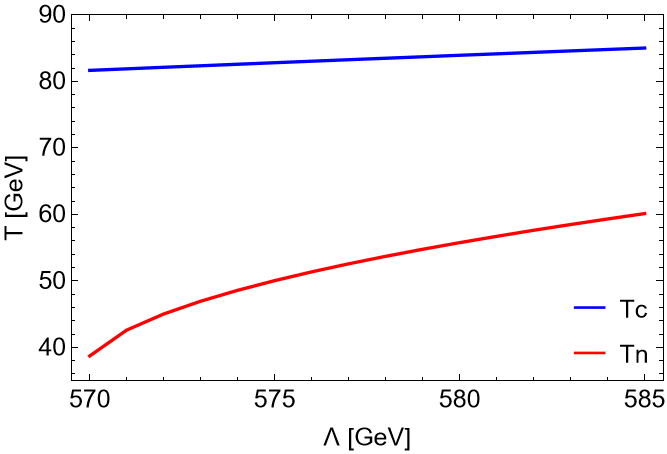}
        \caption{Critical temperature $T_c$ and nucleation temperature $T_n$ as functions of the NP scale $\Lambda$.}
        \label{temperature}
\end{figure}

\begin{table}
\centering
\caption{Benchmark points used in the numerical analysis.}
\label{tab:benchmark}
\begin{tabular}{l c c l c c}
\toprule
Benchmark & $\Lambda~({\rm GeV})$ & $T~({\rm GeV})$ & Benchmark & $\Lambda~({\rm GeV})$ & $T~({\rm GeV})$\\
\midrule
$BM_1$ & 570 & 38.7 &$BM_4$ & 570 & 43.7\\
$BM_2$ & 575 & 38.7 &$BM_5$ & 570 & 48.7\\
$BM_3$ & 580 & 38.7 & &  &  \\
\bottomrule
\end{tabular}
\end{table}

\begin{figure}[!htp]
    \centering
    \begin{minipage}[t]{0.47\textwidth}
        \centering
        \includegraphics[width=\textwidth]{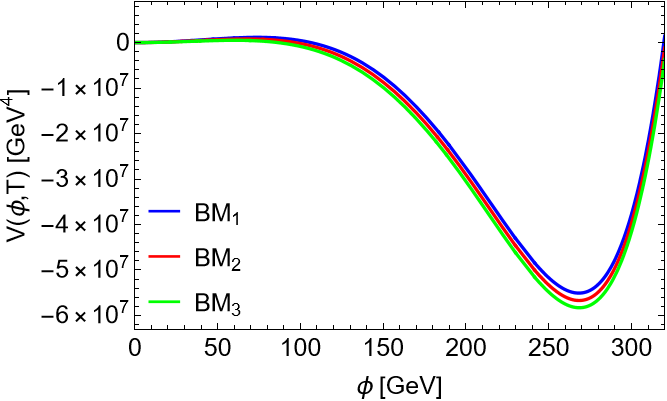}
\caption{Effective potential $V(\phi,T)$ as a function of $\phi$ at fixed temperature $T=38.7~\mathrm{GeV}$ for different values of $\Lambda$. The benchmark parameters are listed in Table~\ref{tab:benchmark}.}
        \label{potential_sameT}
    \end{minipage}
    \hfill
    \begin{minipage}[t]{0.47\textwidth}
        \centering
        \includegraphics[width=\textwidth]{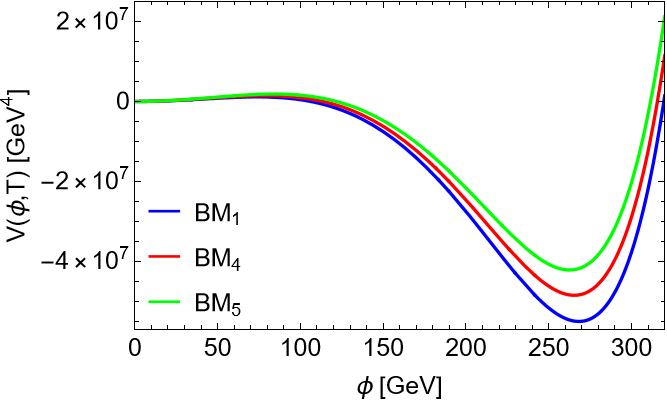}
\caption{Effective potential $V(\phi,T)$ as a function of $\phi$ at fixed NP scale $\Lambda=570~\mathrm{GeV}$ for different temperatures. The benchmark parameters are listed in Table~\ref{tab:benchmark}.}
        \label{potential_samelambda}
    \end{minipage}
\end{figure}

Using the finite-temperature potential specified above, we compute the critical temperature $T_c$ and the nucleation temperature $T_n$ as functions of the NP scale in flat spacetime. The results are shown in Fig.~\ref{temperature}. Over the parameter range considered, both $T_c$ and $T_n$ increase with $\Lambda$.

To illustrate the dependence of the effective potential on $\Lambda$ and $T$, we choose the benchmark points listed in Table~\ref{tab:benchmark}. We take $\Lambda=570~\mathrm{GeV}$ as the reference value, close to the lower boundary of the parameter region admitting successful nucleation in our analysis. This benchmark also gives a GW signal within the sensitivity range \cite{Liu:2025ipj}. The corresponding flat-space nucleation temperature is $T_n\simeq38.7~\mathrm{GeV}$. We vary $\Lambda$ at fixed $T=38.7~\mathrm{GeV}$ to examine the dependence on the NP scale, and vary $T$ at fixed $\Lambda=570~\mathrm{GeV}$ to study the temperature dependence. Fig.~\ref{potential_sameT} shows the finite-temperature potential $V(\phi,T)$ for several values of $\Lambda$ at fixed temperature $T=38.7~\mathrm{GeV}$. From top to bottom, the curves correspond to $\Lambda=570$, $575$, and $580~\mathrm{GeV}$, respectively. The broken-phase minimum shifts slightly toward larger field values as $\Lambda$ increases, from $\phi_{\rm min}=268.16~\mathrm{GeV}$ at $\Lambda=570~\mathrm{GeV}$ to $\phi_{\rm min}=268.45~\mathrm{GeV}$ at $\Lambda=580~\mathrm{GeV}$. Figure~\ref{potential_samelambda} shows the corresponding temperature dependence at fixed $\Lambda=570~\mathrm{GeV}$. As the temperature increases, the broken-phase minimum shifts toward smaller field values.

\begin{figure}[!htp]
    \centering
    \begin{minipage}[t]{0.47\textwidth}
        \centering
        \includegraphics[width=\textwidth]{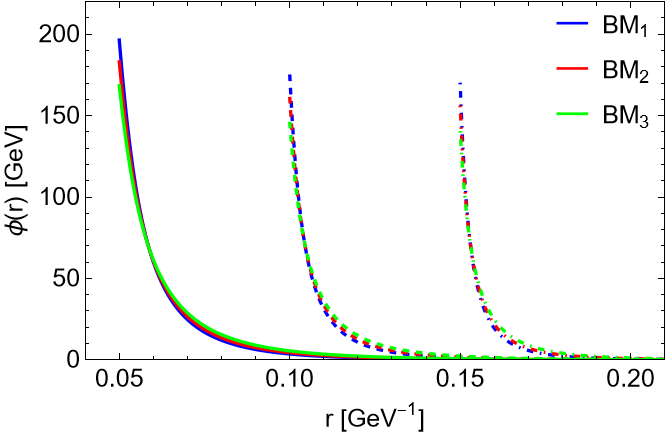}
\caption{Higgs field profiles outside a black hole for several horizon radii, $R_h=0.05$, $0.10$, and $0.15~\mathrm{GeV^{-1}}$, and different values of $\Lambda$ at fixed temperature $T=38.7~\mathrm{GeV}$. The benchmark parameter choices are listed in Table~\ref{tab:benchmark}.}
        \label{configuration_samet}
    \end{minipage}
    \hfill
    \begin{minipage}[t]{0.47\textwidth}
        \centering
        \includegraphics[width=\textwidth]{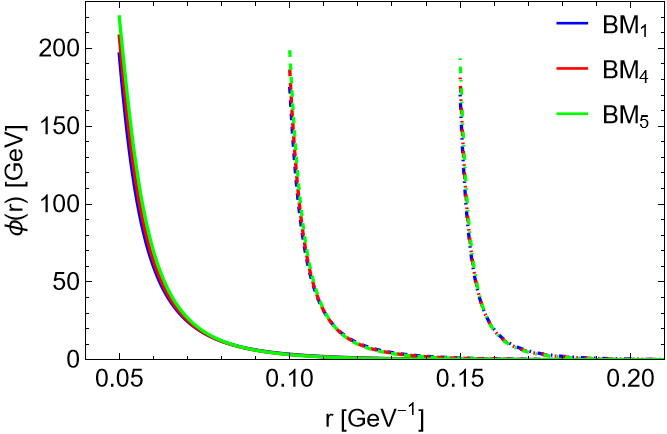}
   \caption{Higgs field profiles outside a black hole for several horizon radii, $R_h=0.05$, $0.10$, and $0.15~\mathrm{GeV^{-1}}$, and different temperatures at fixed NP scale $\Lambda=570~\mathrm{GeV}$. The benchmark parameter choices are listed in Table~\ref{tab:benchmark}.}
        \label{configuration_samelambda}
    \end{minipage}
\end{figure}

\begin{figure}[!htp]
    \centering
    \begin{minipage}[t]{0.47\textwidth}
        \centering
        \includegraphics[width=\textwidth]{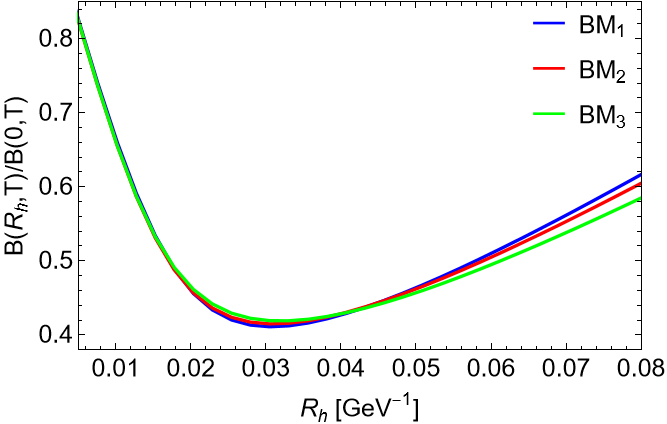}
         \caption{Exponent ratio $\frac{B(R_h,T)}{B(0,T)}$ as a function of $R_h$ at fixed temperature $T=38.7~{\rm GeV}$ for different values of $\Lambda$. The detailed benchmark parameters are listed in Table~\ref{tab:benchmark}.}
        \label{action_ratio_samet}
    \end{minipage}
    \hfill
    \begin{minipage}[t]{0.47\textwidth}
        \centering
        \includegraphics[width=\textwidth]{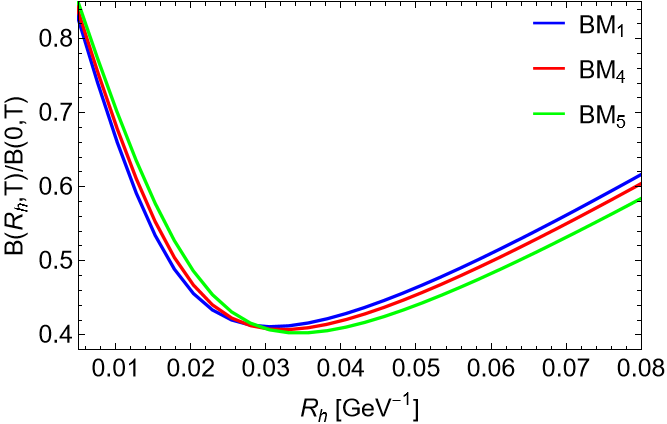}
        \caption{Exponent ratio $\frac{B(R_h,T)}{B(0,T)}$ as a function of $R_h$ at fixed NP scale $\Lambda=570~{\rm GeV}$ for different temperatures. The detailed benchmark parameters are listed in Table~\ref{tab:benchmark}.}
        \label{action_ratio_samelambda}
    \end{minipage}
\end{figure}
Solving the field equation with the corresponding boundary conditions yields the Higgs profiles in the black hole background, as shown in Figs.~\ref{configuration_samet} and~\ref{configuration_samelambda}. At fixed $T=38.7~\mathrm{GeV}$, the horizon value $\phi(R_h)$ decreases with increasing $\Lambda$ for a given $R_h$, whereas it increases with temperature at fixed $\Lambda=570~\mathrm{GeV}$. The horizon value $\phi(R_h)$ need not follow the behavior of the flat spacetime broken-phase minimum $\phi_{\rm min}$. While $\phi_{\rm min}$ is determined locally by $\partial V/\partial\phi=0$, $\phi(R_h)$ follows from the full boundary-value problem in the black hole geometry, including the horizon regularity condition and the false-vacuum boundary condition at spatial infinity.

To quantify the effect of the black hole background on the decay exponent, we plot the ratio $\frac{B(R_h,T)}{B(0,T)}$ in Figs.~\ref{action_ratio_samet} and~\ref{action_ratio_samelambda}. The denominator $B(0,T)$ corresponds to the flat spacetime limit and is used for normalization. Since the thermal decay rate scales as $\Gamma\propto e^{-B}$, $\frac{B(R_h,T)}{B(0,T)} <1$ implies that the black hole weakens the exponential suppression of vacuum decay relative to flat spacetime. Fig.~\ref{action_ratio_samet} shows the decay exponent ratio at fixed temperature $T=38.7~\mathrm{GeV}$ for different values of $\Lambda$. As $\Lambda$ increases, the minimum of $\frac{B(R_h,T)}{B(0,T)}$ increases, while the positions of these minima remain very close. Fig.~\ref{action_ratio_samelambda} shows the corresponding results at a fixed NP scale $\Lambda=570~\mathrm{GeV}$ for different temperatures. Increasing the temperature shifts the location of the minimum toward larger values of $R_h$, while the minimum value of the decay exponent ratio decreases.

The horizon radius is related to the black hole mass $M_{\rm BH}$. In the numerical results presented above, a  modification of the decay exponent occurs for horizon radii corresponding to PBH masses of order $10^{11}$--$10^{13}~\mathrm{g}$. This range arises because the gravitational modification of the critical bubble becomes significant when the horizon radius is comparable to the characteristic radial scale of the Higgs field configuration. We therefore focus on PBHs in this approximate mass range in the following sections.

\section{Gravitational wave production}\label{sec:gw}
\subsection{PBH abundance and the seeded-nucleation condition}
In a radiation-dominated Universe, the mass of a PBH formed at time $t_{\rm f}$ is conventionally parameterized as a fraction $\gamma$ of the horizon mass $M_H$ \cite{carr2021constraints}:
\begin{equation}
     M_{\rm PBH}= \gamma M_{H} \sim 10^{15} \frac{t_{\rm f}}{10^{-23}~{\rm s}} ~{\rm g}.
\end{equation}
The corresponding formation temperature is
\begin{equation}
T_{\rm f}(M_{\rm PBH})=\frac{\sqrt{3}  5^{\frac{1}{4}}}{2 \pi^{\frac{3}{4}}} \sqrt{\gamma} \frac{1}{g_{*}^{\frac{1}{4}}(T_{\rm f})}   \frac{M_{\rm Pl}^{\frac{3}{2}}}{\sqrt{M_{\rm PBH}}}.
\label{formationT}
\end{equation}
Here, $g_{*}(T)$ denotes the effective number of relativistic degrees of freedom contributing to the radiation energy density. The lifetime of a PBH is
\begin{equation}
     \tau= \frac{10240 \pi}{g_{*}(T_{\rm PBH})} \frac{M_{\rm PBH}^3}{M_{\rm Pl}^4}.
\end{equation}
We assume a monochromatic PBH mass function and consider parameter regions in which the PBH lifetime is much longer than the interval between PBH formation and the phase transition. Hawking evaporation is therefore negligible during this period, and the PBH mass can be treated as approximately constant. In addition, the Hawking temperature of considered PBHs is below the ambient radiation temperature during the phase transition. We therefore take the temperature entering the finite-temperature effective potential to be the ambient radiation temperature.

The standard BBN constraint imposes an upper bound on the initial PBH mass fraction at formation, which can be parameterized as \cite{de2021standard}
\begin{equation}
\beta^{\rm PBH}_{1}(T_{\rm f})=\frac{\rho_{\rm PBH}(T_{\rm f})}{\rho_{\rm rad}(T_{\rm f})}\simeq 10^{-19} (\frac{M_{\rm PBH} }{5 \times 10^{-22}M_\odot})^{\frac{1}{2}} f_{\rm PBH},
\label{abundance1}
\end{equation}
For PBHs that survive to the present epoch, $f_{\rm PBH}\equiv\Omega_{\rm PBH}/\Omega_{\rm DM}$ represents their present-day dark-matter fraction. The PBHs considered here, with $10^{11}~{\rm g}\leq M_{\rm PBH}\leq10^{13}~{\rm g}$, have already evaporated, so we use $f_{\rm PBH}$ only as a convenient parameterization of the initial abundance. The corresponding BBN bound is approximately $f_{\rm PBH}\lesssim10^{-4}$ \cite{carr2021constraints}. In this parameter region, PBHs remain subdominant in the total energy density and do not appreciably modify the background thermal evolution. We therefore assume the standard radiation-dominated expansion when evaluating the PBH number density at the phase-transition temperature.

Assuming adiabatic expansion during radiation domination and negligible PBH mass loss, the ratio $\frac{n_{\rm PBH}}{s}$ is conserved. Using $\frac{\rho_{\rm rad}}{s}=\frac{3}{4}\frac{g_*(T)}{g_{*s}(T)}T$ with $g_{*}(T_{\rm f})\simeq g_{*s}(T_{\rm f})$, the number density of considered PBHs at temperature $T$ is
\begin{equation}
    n_{\rm PBH}(T)= \beta^{\rm PBH}(T_{\rm f})  \frac{3 T_{\rm f}}{4 M_{\rm PBH}} \frac{2 \pi^2}{45} g_{*s}(T) T^{3}.
\end{equation}
To obtain the maximum PBH number density allowed by BBN, we set $\beta^{\rm PBH}=\beta_{1}^{\rm PBH}$.

We next examine whether the PBH abundance is sufficient to support PBH-seeded nucleation by comparing the PBH number density with the characteristic bubble number density, $n_{\rm bubble}=\frac{\beta^3}{8\pi v_w^3}$ \cite{hindmarsh2021phase},
where $\beta$ denotes the inverse duration parameter defined below. Assuming that each PBH can seed at most one expanding bubble, we require $n_{\rm PBH}\gtrsim n_{\rm bubble}$. For the subsequent GW analysis, we adopt the reference value $\Lambda=570~{\rm GeV}$ introduced in the previous section. For this choice, we select $R_h=0.0411~{\rm GeV}^{-1}$ as a representative point for which the PBH abundance satisfies $n_{\rm PBH}\gtrsim n_{\rm bubble}$.

For the PBH masses and abundances considered here, the characteristic binary formation time, $t_{bf} \simeq 4 \frac{t_{f}}{(\beta^{\rm PBH}_{1})^{2}}$ \cite{Aljazaeri:2025ftv}, is much longer than the evaporation time. We therefore neglect GW emission from PBH binaries and focus on the GW signal generated by the PBH-seeded phase transition.\footnote{Rare close PBH pairs arising from the spatial distribution, as well as initially clustered PBH populations, would require a separate statistical treatment of binary formation and merger rates and are beyond the scope of the present work. }

\needspace{3\baselineskip}
\subsection{Gravitational wave spectrum}
Because the nucleation temperature in a black hole background remains subject to theoretical uncertainties, we adopt the flat spacetime nucleation temperature as a common reference for the flat and PBH-seeded cases. The inverse duration parameter is then evaluated as
\begin{equation}
    \frac{\beta}{H_n}=T_{n}\frac{dB(R_h, T)}{dT}\Big|_{T=T_{n}}.
\end{equation}
The bubble-wall velocity is estimated as
\begin{equation}\label{vw}
v_w=\begin{cases}
    \sqrt{\frac{\Delta V}{\alpha \rho_{\rm rad}}}&, \quad \sqrt{\frac{\Delta V}{\alpha \rho_{\rm rad}}}<v_{J}(\alpha)\\
    1&,\quad \sqrt{\frac{\Delta V}{\alpha \rho_{\rm rad}}}\geq v_{J}(\alpha),
\end{cases}    
\end{equation}
where $\Delta V=V_{\rm fv}-V_{\rm EW}$ denotes the free-energy difference between the false vacuum and the electroweak vacuum, and $v_{J}(\alpha)=\frac{1}{\sqrt{3}}\frac{1+\sqrt{3\alpha^2+2\alpha}}{1+\alpha}$ is the Jouguet velocity \cite{lewicki2022electroweak}. The strength parameter of the phase transition is defined as 
\begin{equation}
\alpha=\frac{\Delta\rho}{\rho_{\rm rad}}.
\end{equation}
The released energy density is given by
\begin{equation}
\Delta\rho=\left[- \left( V_{\rm EW}-V_{\rm fv} \right)+T \left( \frac{dV_{\rm EW}}{dT}-\frac{dV_{\rm fv}}{dT} \right) \right] \bigg|_{T=T_{n}}.
\end{equation}

With these phase transition parameters, the GW spectrum can be evaluated. In this work, we include contributions from bubble collisions and sound waves. The bubble-wall collision contribution $\Omega_{col}$ can be expressed as \cite{kamionkowski1994gravitational,huber2008gravitational,caprini2016science} 
\begin{eqnarray}
 \label{co}
\Omega_{ col} h^2(f)&\simeq& 1.67\times 10^{-5}\left(\frac{\beta}{H_n}\right)^{-2}\left(\frac{\kappa_\phi\alpha}{1+\alpha}\right)^2\left(\frac{100}{g_*}\right)^{1/3} \notag\\
&\times&\frac{0.11v_w^3}{0.42+v_w^2}\frac{3.8(f/f_{col})^{2.8}}{1+2.8(f/f_{col})^{3.8}}.
\end{eqnarray}
The peak frequency of the collision contribution and efficiency factor $\kappa_\phi$ are
\begin{align}
f_{\rm col}
&= 1.65\times 10^{-5}
\frac{\beta}{H_n}
\left(\frac{0.62}{1.8-0.1v_w+v_w^2}\right)
\nonumber\\
&\quad\times
\left(\frac{T}{100~{\rm GeV}}\right)
\left(\frac{g_*}{100}\right)^{1/6}
~{\rm Hz},
\\
\kappa_\phi
&\simeq
\frac{
0.715\alpha
+\frac{4}{27}\sqrt{\frac{3\alpha}{2}}
}{
1+0.715\alpha
}.
\end{align}

The sound-wave contribution is \cite{hindmarsh2014gravitational,hindmarsh2015numerical,hindmarsh2017shape}
\begin{align}\label{sw}
\Omega_{\rm sw} h^2(f)&=2.65\times 10^{-6}(H_n \tau_{\rm sw})\left(\frac{\beta}{H_n}\right)^{-1}v_w\left(\frac{\kappa_v \alpha}{1+\alpha}\right)^2 \notag \\
&\times\left(\frac{g_*}{100}\right)^{-1/3}\left(\frac{f}{f_{\rm sw}}\right)^3\left(\frac{7}{4+3(f/f_{\rm sw})^2}\right)^{7/2},
\end{align}
where the peak frequency is given by
\begin{equation}
 f_{\rm sw}=1.9\times 10^{-5}\frac{\beta}{H_n}\frac{1}{v_w}\frac{T}{100\,{\rm GeV}}\left(\frac{g_*}{100}\right)^{\frac{1}{6}}~{\rm Hz}.   
\end{equation}
The lifetime of the sound-wave source is modeled as
\begin{equation}
\tau_{\rm sw}=\min\left[\frac{1}{H_n},\frac{R_*}{\overline{U}_f}\right],
\end{equation}
where $H_{n} R_*=v_w(8\pi)^{1/3}(\frac{\beta}{H_{n}})^{-1}$, and the root-mean-square fluid velocity is approximated by \cite{ellis2019gravitational,caprini2020detecting}
\begin{equation}   
\overline{U}_f^2\approx\frac{3}{4}\frac{\kappa_v \alpha}{1+\alpha}.
\end{equation}
The factor $\kappa_v$ is given by \cite{espinosa2010energy}
\begin{equation}
\kappa_v=\frac{\sqrt{\alpha}}{0.135+\sqrt{0.98+\alpha}}.
\end{equation}
Combining the bubble-wall collision and sound-wave contributions, we obtain the predicted GW spectrum as 
\begin{equation}\label{gw}
\Omega_{\rm GW}h^2=\Omega_{\rm col} h^2(f)+\Omega_{\rm sw} h^2(f).
\end{equation}

\begin{figure}[!htp]
        \centering
        \includegraphics[width=0.47\textwidth]{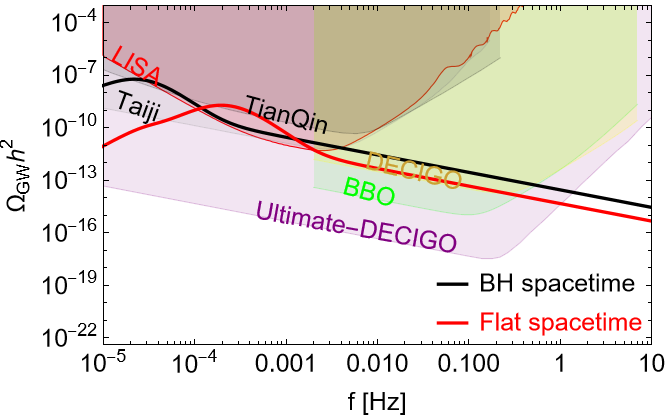}
\caption{GW spectra generated by the first order phase transition in flat spacetime and in a PBH background with $R_h=0.0411~{\rm GeV}^{-1}$, for a fixed NP scale $\Lambda=570~{\rm GeV}$.}
        \label{Gws}
    \end{figure}

Fig.~\ref{Gws} compares the GW spectra in flat spacetime and in the black hole background. In the PBH-seeded case, the peak amplitude is enhanced and the peak frequency is shifted toward lower values. Specifically, the peak frequencies lie in the range $10^{-5}-10^{-4}~{\rm Hz}$ in the black hole background, compared with $10^{-4}-10^{-3}~{\rm Hz}$ in flat spacetime. These modifications are mainly driven by the smaller value of $\frac{\beta}{H_n}$, which corresponds to a longer characteristic phase transition timescale and therefore shifts the characteristic GW frequency to lower values. Parts of the predicted spectra lie above the projected sensitivity curves of LISA \cite{Klein:2015hvg}, TianQin \cite{TianQin:2015yph}, Taiji \cite{Ruan:2018tsw}, and Ultimate DECIGO \cite{Musha:2017usi}, suggesting that these signals may be detectable by future space-based interferometers.

Recent studies have considered different prescriptions for incorporating PBH-induced nucleation into the phase-transition dynamics, showing that the resulting GW spectrum can depend sensitively on how the homogeneous and PBH-induced nucleation channels are treated \cite{Zhong:2025xwm,Tanaka:2026geo}. Other work has further shown that local black hole effects, including the deformation of off-center bubbles, may provide additional modifications to the GW signal \cite{Chang:2025rda}. These approaches differ from the prescription adopted here, where we use the flat spacetime nucleation temperature as a common reference and evaluate the phase-transition parameters using the PBH-modified action.

\section{Sphaleron energy}\label{sec:sphaleron}
 The electroweak $SU(2)_L$ gauge-Higgs sector possesses a topologically nontrivial vacuum structure, with degenerate adjacent vacua separated by an energy barrier whose saddle-point configuration is the sphaleron \cite{manton1983topology,klinkhamer1984saddle}. The thermally activated $B+L$-violating processes in the broken electroweak phase would be exponentially suppressed by sphaleron energy \cite{dashen1974nonperturbative,forgacs1984topology,burzlaff1984classical,yaffe1989static,riotto1999recent}. 
We focus on the Higgs doublet field $\Phi$ along with the $SU(2)_L$ gauge field $W_{\mu}^{a}$. The $U(1)_Y$ hypercharge gauge group is neglected, as its inclusion does not qualitatively affect the results. To investigate the effect of a black hole background on sphaleron configurations, we start from the action \cite{de2021standard}
\begin{align}
S &=\int_{\mathcal{M}}\mathrm{d}^{4}x\sqrt{-g}[\frac{\mathcal{R}}{16\pi G}-g^{\mu\nu}(D_{\mu}\Phi)^{\dagger}D_{\nu}\Phi-\tilde{V}_{\rm eff}(\phi,T)\notag\\
&-\frac{1}{4}g^{\mu\rho}g^{\nu\sigma}F_{\rho\sigma}^{a}F_{\mu\nu}^{a}]+\frac{1}{8\pi G}\int_{\partial\mathcal{M}}K\sqrt{\gamma}\mathrm{d}^{3}y,
\end{align}
where
\begin{align}
F_{\mu\nu}^{a}&=\partial_{\mu} W_{\nu}^{a}-\partial_{\nu} W_{\mu}^{a}+g_2 \epsilon^{a b c} W_{\mu}^{b} W_{\nu}^{c}, \\
D_{\mu} \Phi&=\partial_{\mu} \Phi-\frac{1}{2} i g_2 \sigma^{a} W_{\mu}^{a} \Phi,\\
\tilde{V}_{\rm eff}(\phi,T)&=V(\phi,T)-V(v_{t},T).
\end{align}
Here $v_t$ denotes the Higgs field value at the broken-phase minimum at temperature $T$ and $\phi=\sqrt{2\Phi^{\dagger}\Phi}$. By construction, $\tilde{V}_{\rm eff}(v_t,T)=0$, so that the free-energy density at the broken-phase minimum is set to zero. We take the spacetime geometry around the black hole to be static and spherically symmetric. The metric is written in the Schwarzschild-like form given in Eq.~\eqref{sch1}. For the $SU(2)_L$ gauge field and the Higgs doublet, we adopt the sphaleron ansatz \cite{klinkhamer1984saddle},
\begin{align}
W_i^a\sigma^a dx^i
&=
-\frac{2i}{g_2}
f(g_2 v_t r)\,
dU^\infty (U^\infty)^{-1},\\
\Phi
&=
\frac{v_t}{\sqrt{2}}\,
h(g_2 v_t r)\,
U^\infty
\begin{pmatrix}
0\\
1
\end{pmatrix},
\end{align}
where
\begin{equation}
U^\infty
=
\frac{1}{r}
\begin{pmatrix}
z & x+iy\\
-x+iy & z
\end{pmatrix}, 
\qquad
r=\sqrt{x^2+y^2+z^2}.
\end{equation}
Introducing the dimensionless variables $\xi=g_2v_t r$ and $\xi_h=g_2v_tR_h$, we neglect the backreaction of the sphaleron fields when solving for the sphaleron energy and take the background geometry to be Schwarzschild, with $N(\xi)=1-\xi_h/\xi$ and $\delta=0$. The contribution of the sphaleron configuration to the mass is evaluated separately as a perturbative correction. The radial equations of motion are then
\begin{align}
f''(\xi)
&+\frac{\xi_h}{\xi(\xi-\xi_h)}f'(\xi)
\nonumber\\
&=
\frac{2f(\xi)\bigl[1-f(\xi)\bigr]\bigl[1-2f(\xi)\bigr]}
{\xi(\xi-\xi_h)}
-\frac{\xi h^2(\xi)\bigl[1-f(\xi)\bigr]}
{4(\xi-\xi_h)},
\\[1ex]
h''(\xi)
&+\frac{2\xi-\xi_h}{\xi(\xi-\xi_h)}h'(\xi)
\nonumber\\
&=
\frac{2\bigl[1-f(\xi)\bigr]^2h(\xi)}
{\xi(\xi-\xi_h)}
+
\frac{\xi}{(\xi-\xi_h)g_2^2v_t^4}
\frac{\partial \tilde{V}_{\rm eff}(v_t h(\xi),T)}{\partial h}.
\end{align}
The boundary conditions are \cite{de2021standard}
\begin{align}
f(\xi_h)&=0,\qquad h(\xi_h)=0, \\
\lim_{\xi\to\infty}f(\xi)&=1,\qquad
\lim_{\xi\to\infty}h(\xi)=1.
\end{align}

To evaluate the sphaleron energy, we write the mass function as $M(\xi)=M_{\rm BH}+\delta M(\xi)$ with $\delta M(\xi_{h})=0$. The asymptotic mass correction is identified with the sphaleron energy, $E_{sph}(\xi_h, T)= \delta M(\infty)$ \cite{de2021standard}, yielding
\begin{align}
E_{sph}(\xi_h, T)&= \frac{4 \pi v_{t}}{g_2} \int_{\xi_h}^{\infty} \mathrm{d} \xi[ \frac{1}{2}\xi(\xi -\xi_h)h'(\xi)^2 \notag\\
&+\frac{4}{\xi}(\xi -\xi_h)f'(\xi)^2+\frac{8}{\xi^2}f(\xi)^2(1-f(\xi))^2\notag\\
&+h(\xi)^2(1-f(\xi))^2+\frac{\xi^2}{{g_2}^2 v_{t}^{4}} \tilde{V}_{\rm eff}(h(\xi) v_t,T)].
\end{align}
In the limit $\xi_h \to 0$, the metric function $N(r)=1- \xi_h/\xi$ approaches unity and this expression reduces to the usual sphaleron energy in flat spacetime. 
\begin{figure}[!htp]
    \centering
\includegraphics[width=0.45\textwidth]{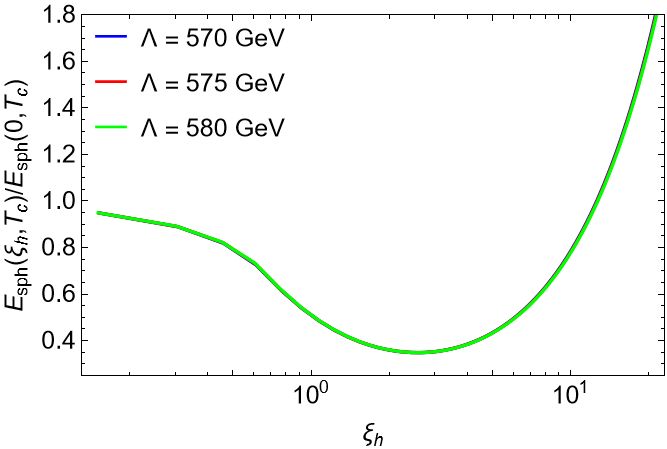}
\caption{Ratio of the sphaleron energy in a black hole background to the corresponding flat spacetime value as a function of the dimensionless black hole horizon radius $\xi_h$ for $\Lambda=570,575,580~{\rm GeV}$. Each curve is evaluated at the critical temperature $T_c$ for the corresponding value of $\Lambda$.}
        \label{sphaleron}
\end{figure}

Fig.~\ref{sphaleron} shows the normalized sphaleron energy $E_{sph}(\xi_{h},T_c)/E_{sph}(0,T_c)$ for different values of $\Lambda$. A ratio below unity indicates that the black hole background lowers the sphaleron energy barrier relative to flat spacetime. Since the finite-temperature sphaleron transition rate is exponentially suppressed as $\Gamma_{sph}\propto e^{- \frac{E_{sph}}{T}}$, a reduction in this ratio corresponds to a weaker exponential suppression. The dimensionless horizon radius is related to the black hole mass through $\xi_{h}=g_{2} v_{t} R_{h}$ and $M_{\rm BH}=R_h/(2G)$, where $v_{t} \equiv v(T_{c} )$. For the parameters considered here, the sphaleron energy is reduced for PBH masses in the approximate range $10^{11}~{\rm g}-10^{13}~{\rm g}$. This behavior reflects the fact that the black hole effect is strongest when the horizon radius becomes comparable to the characteristic sphaleron size. At finite temperature, the black hole mass range associated with the strongest reduction in the sphaleron energy is shifted toward larger masses relative to the zero-temperature estimate based on $R_h \sim \frac{1}{g_2 v_{0}}$ \cite{de2021standard}. This shift arises because $v_{t}<v_{0}=246~{\rm GeV}$, resulting in a larger characteristic sphaleron length scale $\frac{1}{g_2 v_{t}}$.

Using the BBN upper bound on the PBH abundance given above, the corresponding PBH-to-baryon number ratio can be estimated as
\begin{equation}
\frac{n_{\rm PBH}}{n_{\rm b}}\sim 10^{-34}\left(\frac{ \eta}{10^{-9}}\right)^{-1}\left(\frac{M_{\rm PBH}}{5 \times 10^{-22}M_{\odot}}\right)^{-1}f_{\rm PBH}.
\end{equation}
Although the black hole background lowers the sphaleron energy and thereby reduces the exponential suppression of local sphaleron transitions, the PBH-to-baryon number ratio is estimated to satisfy $\frac{n_{\rm PBH}}{n_{\rm B}} \lesssim 10^{-37}$ in the mass range considered here. Following the order-of-magnitude argument of Ref.~\cite{de2021standard}, this small PBH-to-baryon number density ratio suggests that PBH-seeded sphaleron transitions are unlikely to appreciably wash out a pre-existing cosmological baryon asymmetry.

The above discussion concerns the gravitational modification of the sphaleron configuration by the black hole background. A distinct role of sphaleron processes can arise in the hot spots generated by PBH evaporation. In such regions, sphaleron processes can remain locally active after their freeze-out in the background Universe, allowing the lepton asymmetry generated by PBH-emitted right handed neutrinos to be converted into baryon asymmetry \cite{Gunn:2024xaq}. For PBH in the masses range $10^{11}~{\rm g}<M_{\rm PBH}<10^{13}~{\rm g}$, the initial hotspot core temperature is far below $T_{\rm sph}\simeq130~{\rm GeV}$, and the local sphaleron active regime can only be reached during the late stages of evaporation, when the instantaneous PBH mass has decreased to approximately $10^7~{\rm g}$. In this mass range, the hotspot dynamics can also modify the BBN constraints on the PBH abundance, since the locally heated plasma can suppress the escape of low-energy photons and thereby relaxing the BBN photo-dissociation constraints \cite{Altomonte:2025hpt}. A consistent assessment of hotspot-assisted leptogenesis for $10^{11}~{\rm g}<M_{\rm PBH}<10^{13}~{\rm g}$ PBHs would therefore require a simultaneous treatment of the late-time evaporation history, right-handed-neutrino production and decay, the evolving hotspot profile, and the corresponding hotspot-modified BBN constraints. We leave such an analysis for future work.

\section{Baryogenesis via Hawking radiation} \label{sec:baryon}
We next consider a different PBH-induced baryogenesis mechanism that is independent of the nucleation temperature discussed above. In this mechanism, BSM particles $X$ produced through Hawking evaporation generate a net baryon asymmetry directly through their subsequent baryon-number-violating and CP-violating decays. Provided that the resulting asymmetry is not erased by washout processes, these decays can contribute to the cosmological baryon asymmetry. Neglecting greybody factors, the differential particle-number emission rate for $X$ from a Schwarzschild black hole can be approximated as \cite{Hawking:1975vcx}
\begin{equation}
\frac{d^2 N_X}{dt\,dE}=\frac{4\pi r_h^2}{E}\frac{d^2 u_X}{dt\,dE}=\frac{g_X}{2 \pi}\frac{r_h^2 E^2}{e^{E/T_{\mathrm{\rm PBH}}}\pm 1}.
\end{equation}
Here, the $+$ and $-$ signs correspond to fermions and bosons, respectively. In the numerical analysis below, we take $X$ to be a bosonic with $g_X=1$. The total number of $X$ particles emitted during the lifetime of the black hole is obtained by integrating the differential emission rate over the particle energy and the full evaporation history. For bosons, this integration yields \cite{Gondolo:2020uqv}
\begin{align}
N_X &=\frac{120 \zeta(3)}{\pi^3}
\frac{g_{\rm X}}{g_*(T_{\rm PBH})}
\frac{M_{\rm PBH}^2}{M_{\mathrm{Pl}}^2},
&\quad
T_{\rm PBH}>m_{\rm X}, \\
N_X &=\frac{15 \zeta(3)}{8\pi^5}
\frac{g_{\rm X}}{g_*(T_{\rm PBH})}
\frac{M_{\rm Pl}^2}{m_{\rm X}^2},
&\quad
T_{\rm PBH}<m_{\rm X}.
\end{align}
Here, the $T_{\rm PBH}$ appearing in the two emission regimes denotes the initial Hawking temperature associated with the initial PBH mass $M_{\rm PBH}$. Since we only aim at an order-of-magnitude estimate, we neglect the variation of $g_*(T_{\rm PBH})$ and adopt a fixed benchmark value $g_*(T_{\rm PBH})=106.8$ throughout this section. For the PBH masses considered here, this approximation affects the emitted particle number only at the $\mathcal O(1)$ level. For completeness, the corresponding fermionic result is related to the bosonic one by $N_{F}=\frac{3}{4} \frac{g_{F}}{g_{B}}N_{B}$, where $g_F$ and $g_B$ denote the corresponding numbers of degrees of freedom. 

The initial PBH abundance determines whether PBHs become dominant before they evaporate. The PBH energy density scales as $\rho_{\rm PBH}\propto a^{-3}$, whereas the radiation energy density scales as $\rho_{\rm rad}\propto a^{-4}$. Consequently, the ratio $\rho_{\rm PBH}/\rho_{\rm rad}$ increases with the scale factor. We denote by $T_{\rm early-eq}$ the temperature at which the PBH and radiation energy densities become equal. Using $\rho_{\rm PBH}/\rho_{\rm rad}\propto a\propto T^{-1}$ during radiation domination, we obtain 
\begin{align}
\frac{\rho_{\rm PBH}(T_{\rm early-eq})}
{\rho_{\rm rad}(T_{\rm early-eq})}
&=
\frac{\rho_{\rm PBH}(T_{\rm f})}
{\rho_{\rm rad}(T_{\rm f})}
\frac{T_{\rm f}}{T_{\rm early-eq}}
\notag\\
&=
\beta^{\rm PBH}(T_{\rm f})
\frac{T_{\rm f}}{T_{\rm early-eq}}
\simeq 1.
\end{align}
Here, $T_{\rm f}$ denotes the PBH formation temperature as given in Eq.~(\ref{formationT}). If equality occurs before PBH evaporation, $t_{\rm early-eq}<t_{\rm eva}$, the Universe enters a PBH-dominated phase. The critical abundance $\beta_c^{\rm PBH}$ corresponds to the limiting case $t_{\rm early-eq}=t_{\rm eva}$, or equivalently $T_{\rm early-eq}=T_{\rm eva}$. The temperature of the Universe at the PBH evaporation time is given by
\begin{equation}
T_{\rm eva}
=
\frac{\sqrt{3}}
{64\sqrt{2}\,5^{\frac{1}{4}}\pi^{\frac{5}{4}}}
\frac{g_*^{\frac{1}{2}}(T_{\rm PBH})}
{g_*^{\frac{1}{4}}(T_{\rm eva})}
\frac{M_{\rm Pl}^{\frac{5}{2}}}
{M_{\rm PBH}^{\frac{3}{2}}},
\end{equation}
and the critical initial PBH mass fraction is therefore \cite{Gondolo:2020uqv}
\begin{align}
\beta^{\rm PBH}_c(T_{\rm f})
&=
\frac{T_{\rm eva}}{T_{\rm f}}
=
\sqrt{\frac{g_{*}(T_{\rm PBH})}{10240\pi\gamma}}
\frac{M_{\rm Pl}}{M_{\rm PBH}}
\notag\\
&\simeq
2.8\times10^{-6}
\left(\frac{g_{*}(T_{\rm PBH})}{106.8}\right)^{1/2}
\notag\\
&\times\left(\frac{0.2}{\gamma}\right)^{1/2} \left(\frac{1~{\rm g}}{M_{\rm PBH}}\right),
\end{align}
where we adopt the benchmark value $\gamma=0.2$ for the PBH formation efficiency \cite{Carr:1975qj,Sasaki:2018dmp}. For PBHs in the mass range $10^{11}~{\rm g}<M_{\rm PBH}<10^{13}~{\rm g}$, the corresponding formation temperature is $T_{\rm f}\sim10^{9}-10^{10}~{\rm GeV}$, so that the PBHs form well before the electroweak-scale phase transition. Their evaporation occurs at background temperatures of approximately the ${\rm eV}-{\rm keV}$ scale, well below the characteristic BBN temperature, $T_{\rm BBN}\sim0.1-1~{\rm MeV}$, but above the recombination temperature, $T_{\rm rec}\sim0.26~{\rm eV}$ \cite{ParticleDataGroup:2014cgo}. The PBHs therefore survive throughout BBN and evaporate after BBN but before recombination, corresponding to $T_{\rm f}>T_{n}>T_{\rm BBN}>T_{\rm eva}>T_{\rm rec}$. By contrast, $T_{\rm PBH}$ is the Hawking temperature of a black hole and scales as $T_{\rm PBH}\propto M_{\rm PBH}^{-1}$, so its ordering relative to $T_{n}$ depends on $M_{\rm PBH}$.

To quantify the baryon asymmetry generated by the decays of the PBH-emitted $X$ particles, we introduce the dimensionless CP-asymmetry parameter
\begin{equation}
\gamma_{CP} =\sum_{i} B_{i} \frac{\Gamma (X \to f_{i})- \Gamma (\bar{X} \to \bar{f_{i}})}{\Gamma_{X}}.
\end{equation}
Here, $B_i$ denotes the baryon number carried by the final state $f_i$, and $\Gamma_X$ is the total decay width of $X$. $f_{i}$ and $\bar{f_{i}}$ are a particle and its antiparticle, respectively. The baryon asymmetry is generated through the interference between the tree-level and one-loop decays of $X$. Although the precise value of $\gamma_{CP}$ is model-dependent, CP asymmetries generated through the interference between tree-level and one-loop decays are loop suppressed \cite{Gehrman:2022imk}. Accordingly, we adopt the representative value $\gamma_{CP}=10^{-2}$. 

For a monochromatic PBH population in radiation-dominated regime, the resulting baryon-to-entropy ratio is given by \cite{Gehrman:2022imk}
\begin{align}
Y_{B}
&= \frac{n_{\rm B}(t_0)}{s(t_0)}
 = \gamma_{ CP}
   \frac{n_{\rm X}(t_{\rm eva}+\tau_{\rm X})}
        {s(t_{\rm eva}+\tau_{\rm X})}
 = \gamma_{ CP}
   \frac{n_{\rm X}(t_{\rm eva})}
        {s(t_{\rm eva})} \notag\\
&= \gamma_{ CP} N_{\rm X}
   \frac{n_{\rm PBH}(t_{\rm f})}{s(t_{\rm f})}
 = \gamma_{ CP}\beta^{\rm PBH} N_{\rm X}
   \frac{1}{M_{\rm PBH}}
   \frac{\rho_{\rm rad}(t_{\rm f})}{s(t_{\rm f})} \notag\\
&= \frac{3}{4}
   \frac{g_{*}(T_{\rm f})}{g_{*s}(T_{\rm f})}
   \gamma_{ CP}\beta^{\rm PBH} N_{\rm X}
   \frac{T_{\rm f}(M_{\rm PBH})}{M_{\rm PBH}},
\end{align}
where $\tau_X$ denotes the lifetime of $X$. Under the prompt-decay assumption, the baryon asymmetry is generated close to the PBH evaporation epoch. We further require $m_{\rm X}>T_{\rm eva}$ so that the inverse processes associated with $X$ decay are Boltzmann suppressed, thereby avoiding washout of the generated baryon asymmetry. Separately, PBH evaporation is subject to BBN constraints. For $10^{11}~{\rm g}<M_{\rm PBH}<10^{13}~{\rm g}$, the standard constraint from directly emitted SM particles is given by $\beta^{\rm PBH}_{1}$ in Eq.~\eqref{abundance1}. The subsequent decays of the PBH-emitted $X$ particles may introduce additional BBN constraints, but these depend on the detailed decay time, decay channels, and branching fractions of $X$ \cite{Kawasaki:2017bqm}. Since these properties are model dependent and are not specified here, we impose only $\beta^{\rm PBH}\leq\beta^{\rm PBH}_{1}$. As $Y_B\propto\beta^{\rm PBH}$ in the radiation-dominated regime, setting $\beta^{\rm PBH}=\beta^{\rm PBH}_{1}$ gives a conservative upper estimate of the baryon asymmetry within the assumptions adopted here. A decay-induced BBN constraint would modify this estimate only if it were more stringent than the standard PBH bound.

\begin{figure}[!htp]
    \centering
\includegraphics[width=1\linewidth]{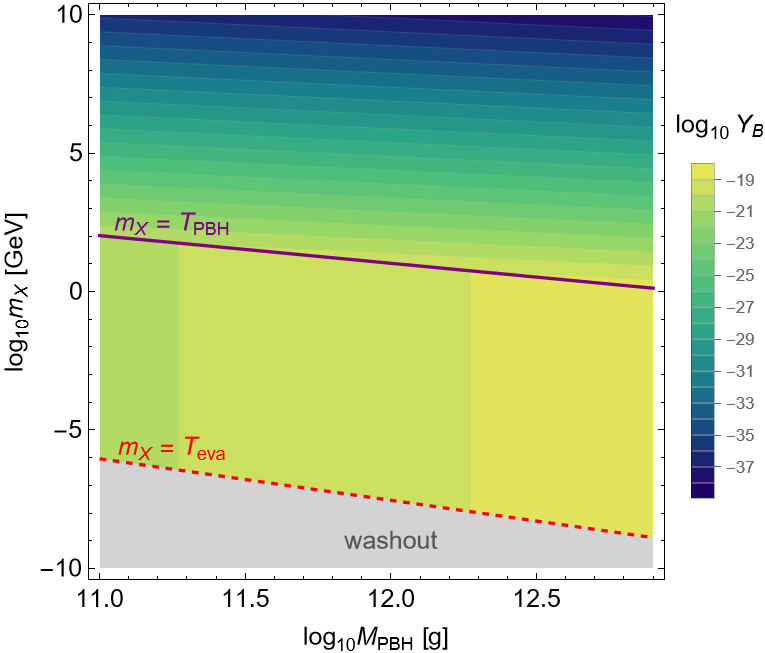}
    \caption{Contours of $\log_{10}Y_{B}$ in the ($\log_{10}m_{X}$, $\log_{10}M_{\rm PBH}$) plane for $\gamma_{CP}=10^{-2}$.}
    \label{baryon}
\end{figure}

We then calculate $Y_B$ and display its contours in Fig.~\ref{baryon}. As a criterion for avoiding thermal washout, we exclude the shaded gray region where $m_X<T_{\rm eva}$. The red dashed and purple curves correspond to $m_{\rm X}=T_{\rm eva}$ and $m_{\rm X}=T_{\rm PBH}$, respectively. Within the parameter region allowed by the adopted BBN and washout criteria, PBHs with $10^{11}~{\rm g} < M_{\rm PBH} < 10^{13}~{\rm g}$ produce $Y_B < 10^{-18}$, corresponding to $\frac{Y_B}{Y_{B}^{obs}} \lesssim 10^{-8}$ with the observational value $Y_{B}^{obs} \simeq 8.7 \times 10^{-11}$ \cite{ParticleDataGroup:2020ssz}. This small baryon yield primarily results from the stringent BBN upper bound on the initial PBH mass fraction. Given the small total baryon asymmetry generated by PBH evaporation under the assumptions adopted here, the additional contribution to the baryon abundance between the BBN and CMB epochs is negligible and does not appreciably affect the consistency between the values inferred from BBN and CMB observations.

\section{Conclusions and Discussions}
\label{sec:conclusion}
In this work, we investigate the effects of PBH backgrounds on the vacuum decay exponent, the GW spectrum, and the sphaleron energy, as well as baryogenesis from BSM particles emitted through Hawking radiation. We find that the decay exponent in the PBH background is smaller than the corresponding flat spacetime value, indicating that PBH can weaken the exponential suppression of the false vacuum decay. At fixed temperature, increasing the NP scale raises the minimum value of the exponent ratio $B(R_h,T)/B(0,T)$, while the positions of these minima remain close to each other. At fixed $\Lambda$, increasing the temperature shifts the minimum of the exponent ratio toward larger PBH masses. For the PBH mass range around $10^{11}~{\rm g}-10^{13}~{\rm g}$ considered in this work, the decay exponent is reduced. 
 
We then analyze the effect of the PBH background on the GW spectrum from the first-order phase transition. We consider PBH-centered, spherically symmetric bubbles, assuming $n_{\rm PBH}\gtrsim n_{\rm bubble}$, and use the flat spacetime nucleation temperature as the reference temperature because the PBH abundance is small and the Hawking temperature is below the ambient temperature. Relative to flat spacetime, the PBH background enhances the GW peak amplitude and shifts the peak frequency to lower values, corresponding to a longer characteristic transition timescale. Parts of the predicted spectra fall within the sensitivity ranges of LISA \cite{Klein:2015hvg}, TianQin \cite{TianQin:2015yph}, Taiji \cite{Ruan:2018tsw}, and Ultimate DECIGO \cite{Musha:2017usi}. Since the resulting GW spectrum depends on how PBH-induced nucleation is incorporated into the phase-transition dynamics, different prescriptions can lead to different modifications of the nucleation history and its dependence on the PBH abundance \cite{Zhong:2025xwm,Tanaka:2026geo}.

We next study the effect of PBHs on the sphaleron energy. For black hole masses of order $10^{11}~{\rm g}-10^{13}~{\rm g}$, the PBH background lowers the sphaleron energy, thereby weakening the exponential suppression of sphaleron transitions. Nevertheless, owing to the small PBH abundance considered in this work, the associated washout effect is not expected to appreciably modify a pre-existing cosmological baryon asymmetry. 
Finally, adopting initial PBH abundances consistent with constraints from primordial light-element abundances, we estimate the baryon asymmetry generated by the Hawking emission of BSM particles $X$ and their subsequent decays and compare it with the observed value. The resulting baryon asymmetry from PBHs in the mass range $10^{11}~{\rm g}-10^{13}~{\rm g}$ is much smaller than the observed baryon asymmetry, with $\frac{Y_B}{Y_{B}^{obs}} \lesssim 10^{-8}$. Consequently, the post-BBN baryon asymmetry generated by PBH evaporation is negligible and does not appreciably affect the agreement between the baryon abundances inferred from BBN and CMB observations.

\normalsize
\section{Acknowledgments}
This work is supported by the National Natural Science Foundation of China (NSFC) under Grants Nos.  12322505, 12547101, and 	1267051668. L.B. also acknowledges Chongqing Natural Science Foundation under Grant
No. CSTB2024NSCQ-JQX0022 and
Chongqing Talents: Exceptional Young Talents Project No. cstc2024ycjh-bgzxm0020.

\end{CJK}

\bibliographystyle{apsrev4-1}
\bibliography{ref}

\end{document}